# Thermodynamics of small systems by nanocalorimetry: from physical to biological nano-objects.

J.-L. Garden*, H. Guillou, A.F. Lopeandia, J. Richard, J-S. Heron, G.M. Souche, F.R. Ong, B. Vianay, O. Bourgeois.

*Institut NÉEL, CNRS et Université Joseph Fourier, BP 166, 38042 Grenoble Cedex 9, France*

**Abstract**

Membrane based nanocalorimeters have been developed for ac calorimetry experiments. It has allowed highly sensitive measurements of heat capacity from solid state physics to complex systems like polymers and proteins. In this article we review what has been developed in ac calorimetry toward the measurement of very small systems. Firstly, at low temperature ac calorimetry using silicon membrane permits the measurement of superconducting sample having geometry down to the nanometer scale. New phase transitions have been found in these nanosystems illustrated by heat capacity jumps versus the applied magnetic field. Secondly, a sensor based on ultra-thin polymer membrane will be presented. It has been devoted to thermal measurements of nanomagnetic systems at intermediate temperature (20K to 300K). Thirdly, three specific polyimide membrane based sensors have been designed for room temperature measurements. One is devoted to phase transitions detection in polymer, the second one to protein folding/unfolding studies and the third one will be used for the study of heat release in living cells. The possibility of measuring systems out of equilibrium will be emphasized.

# 1. Introduction

Since a decade, the strong interest devoted to nanosciences brought the need of developing new thermal sensors. These new devices are also downscaled to be adapted to the measurement of very small mass sample required by the continuous reduction of size in nanoscopic physics. Because calorimetry is a unique tool to characterize the presence of unknown nanophases in modern materials, membrane based nanocalorimeters are a very important tool for thermophysical analysis. This type of calorimeter is perfectly adapted to the highly sensitive measurement of heat capacity at the nanoscale especially if associated with ac calorimetry.

The universal nature of specific heat measurement allows the investigation of such a large field of research like condensed matter, soft matter or even nuclear physics. The application of nanocalorimetry has been done in fundamental researches in condensed matter physics at the nanoscale at low and very low temperature as well as studies on soft matter objects such as polymers at room temperature. Equally, the study of biological macromolecules and microscopic objects such as living cells require the use of highly sensitive thermal sensors.

Our purpose in this article is to give an overview of all the membrane based ac-nanocalorimeters which has been built in our group, each of them being devoted to a precise physical experiment on small systems. We provide technical information on design, electronics and sensitivity of the measurement. For each ac-calorimeter a physical experiment made with it is briefly reported, details are available in previous publications.

One particular calorimetric method used at low temperature is the so-called ac-calorimetry. This method has two special features. Firstly, it is one of the most sensitive in terms of detection of small heat capacity variations, and secondly, the samples under study have to be very small in order to be homogeneous in temperature. The latest technological developments in micro and nanofabrication facilities have enabled the conception and realization of sensors of very low mass and consequently of very low heat capacity addenda. The capabilities in terms of low thermal events detection are particularly enhanced which allow measurements of new nanophases and physical thermal events never observed before. Let us just briefly mention that all along the paper we employ the expression nanocalorimetry because the limit of detection in energy of our device is at least of the order of nanojoule. The term nanojoule-calorimetry could also be valuable, like picojoule-calorimetry around 100K or attojoule-calorimetry below 1K indicating the energy sensitivity of the different devices at their working temperature.

The paper is organized in three sections. After the exposition of the principal features of the steady state ac-calorimetry technique in section 2, the calorimetry at the nanoscale at low temperatures is presented in the section 3. At such low temperatures, it will be emphasized that new mesoscopic phenomena may occur at the interface between the quantum and the macroscopic worlds. In the section 4, soft matter nanocalorimetry at ambient temperature is discussed. The thermodynamic behaviours of complex systems taking place outside equilibrium will be particularly outlined.

**2. Basic principles of the ac-calorimetry method**

In this introductory section, we briefly recall the principle of the ac-calorimetry method used in most of our experiments.

*2.1 Basic equations*

An input thermal power $P(t) = P_{dc} + P_{ac}$ constituted by a dc and an ac term is supplied to a system of well defined temperature $T$ connected to a thermal bath (constant temperature $T_0$ and infinite heat capacity) by means of a thermal conductance $K_b$. In the linear regime assumption, and under stationary conditions, the temperature response is composed by a dc and an ac component:

$$\begin{cases} \Delta T_{dc} = T_{dc} - T_0 = \dfrac{P_0}{K_b} \\ T_{ac}^* = \delta T_{ac} \exp[i(\omega t - \varphi)] \end{cases} \quad (1)$$

The star indicates complex notations of the oscillating variables used here for the sake of simplicity. $T_{dc}$ is the mean constant dc temperature of the sample, $T_0$ is the constant temperature of the bath, $\delta T_{ac}$ is the amplitude of the oscillating temperature, $\omega$ is the angular frequency and $\varphi$ is the phase shift between the oscillating temperature and the input oscillating heat flow with a phase taken by convention equal to zero:

$$P_{ac}^* = P_0 \exp(i\omega t) \quad (2)$$

Experimentally particularly at low frequencies the measurement is not adiabatic or quasi-adiabatic, and at high frequencies the temperature of the sample is not homogeneous anymore. This means that the external relaxation time $\tau_{ext} = C/K_b$ and the diffusion time $\tau_{int} = L^2 / (D\sqrt{90})$ have to be taken into account [1]. $C$ is the total specific heat, $L$ is the thickness of the sample and $D$ is the thermal diffusivity of the sample. In this case, the amplitude of the temperature oscillation takes the following form [1]:

$$\delta T_{ac} = \frac{P_0}{\omega C}\left[1 + \frac{1}{(\omega\tau_{ext})^2} + (\omega\tau_{int})^2 + \frac{2K_b}{3K_s}\right]^{-1/2} \qquad (3)$$

$K_b$ and $K_s$ are the thermal conductance connecting the sample to the thermal bath and the internal thermal conductance inside the volume of the sample respectively. Now we face up to the experimental challenge of designing a calorimetric device with an appropriate working frequency. The frequency of the temperature oscillation has to be faster than the relaxation time to the heat bath (adiabatic condition) and slower than the diffusion time in the sensor. This last condition ensures that the temperature can be considered as homogeneous over the sensor. These two conditions are met when the angular frequency fulfils the following inequalities:

$$(\omega\tau_{int})^2 \ll 1 \ll (\omega\tau_{ext})^2 \qquad (4)$$

In the following we discuss the two important conditions, the quasi-adiabaticity and the homogeneity of the temperature into the sample.

-adiabaticity: $\omega\tau_{ext} \gg 1$

The amount of heat is supplied and taken to the sample over a period which is faster than the external relaxation time ($\tau_{ext}$) of the sample towards the thermal bath. This makes a limitation to the lowest frequency that can be used for meaningful measurements. This condition is by definition not respected for the temperature modulated differential scanning calorimetry (TMDSC), where the heat exchange coefficient linking the sample to the surrounding has to be known [2].

-homogeneity of the sensor temperature: $\omega\tau_{int} \ll 1$

The amount of heat is consecutively supplied and taken to the sample over a period which is larger than the internal relaxation time constant ($\tau_{int}$) of the temperature inside the sample and its cell-holder. This introduces a limitation for the temperature oscillation at high frequency. These two previous conditions are experimentally determined by the measurement of the adiabatic plateau as explained in the following.

Out of this frequency range, the temperature oscillation undergoes a phase shift different from π/2 and the amplitude follows equation (3). In the expression (3), the heat capacity of the thermometer, the heater, and the conductance due to bad thermal contacts between these different elements and the sample have been neglected. Precise calculations of the amplitude and the phase of the temperature oscillation can be found in the classical reference [1]. Experimentally, when we plot $\omega\delta T_{ac}$ versus frequency, in the range of frequency where the inequalities (4) are valid, we obtain what is usually known as the "adiabatic plateau" where $\omega\delta T_{ac}=P_0/C$.

It is however possible, to a certain extent, to correct the measured signal out of the plateau such as described in the next paragraph for particular situations.

*2.2 The dynamic aspect of the ac-calorimetry*

At low temperatures the relaxation times of the dynamic events (quantum effects, phase transitions, etc…) are very small as compared to the period of the temperature oscillation, the heat capacity which is inferred from the ac-calorimetry takes its classical thermodynamic meaning. The measured heat capacity is a quasi-static property. This is not true for slow phase transformations in complex systems. Then the dynamic aspect can be directly probed by the ac-calorimetry method. In this case, the measured heat capacity is a generalized calorimetric susceptibility with a real and an imaginary component having different physical meanings [3-5]:

$$C^* = C' - iC'' = C_\infty + \frac{C_0 - C_\infty}{1 + i\omega\tau} \tag{5}$$

This simple Debye-relaxation relation points out the appearance of a time scale separation between different internal degrees of freedom involved in the phase transition. $C_\infty$ is the heat capacity of the vibrational modes (phonon bath) or other very fast thermalized degrees of freedom (electrons bath, etc...). On the other hand, $C_0 - C_\infty$ is the configurational contribution to the heat capacity of the structural degrees of freedom involved in the phase transition. This contribution results from the very slow thermalization of these modes as compared to the phonons or other fast degrees of freedom. Generally, this latter contribution is called the enthalpy peak because it is related to the isothermal heat of the transformation during the advance of the process. In principle, the separation of time scale between the two heat capacity contributions can be observed when the frequency of the temperature oscillation is varied over several decades. Unfortunately, the operating frequency range is narrow in ac-calorimetry because of the experimental constraints given by the inequalities (4). Nevertheless if we are able to perfectly work out a thermal model for the sample holder and the sample we can extract the heat capacity beyond the frequency range determined by the inequalities (4) [6]. Such a calculation requires an exact knowledge of the addenda. For ac-calorimetry, the extracted data can be precisely linked to the equation (5) if the two following conditions can be controlled:

-stationary condition: $\dot{T}_{dc} \ll \dot{T}_{ac}^*$ (The dot is used for time derivative)

The dc temperature rate (scanning ramp) has to be smaller than the instantaneous ac temperature rate $\dot{T}_{ac}^* = i\omega T_{ac}^*$. This condition is by definition not fulfilled in TMDSC where the influences of the dc and ac dynamic temperature components have to be separated by deconvolution procedure.

-linearity: $\dot{T}_{ac}^* << T_0/\tau$

The derivation of this inequality will be proposed in a forthcoming paper. Here $\tau$ is the relaxation time of the slow physical event of interest, the same used in Eq. (5). Since this condition depends directly on the physics, the frequency and the amplitude of the oscillating temperature have to be chosen in order to fulfil this condition. For very slow processes such as glass transitions it can be rather difficult to remain in the linear regime.

## 3. Low temperature nano-calorimetry

*3.1 Low temperature highly sensitive ac calorimetry*

In this section we present a silicon membrane based calorimeter designated for low temperature measurement of heat capacity. The targeted scientific objectives are the measurement of systems structured down to the nanoscale. It is now well established that the thermodynamic as well as the thermal properties of nanosystems are quite different from the bulk especially at very low temperature. The competition between various characteristic length scales is at the origin of new thermal phenomena. Indeed, surface to volume ratio, effect of the topology or finite size effects perturb and/or control the thermal/thermodynamic properties of downscaled devices [7,8]. We first present the sensor and the measurement method and then we give results obtained on superconducting nanosystems. Finally we show how this sensor allows the measurement of mesoscopic superconductor systems like nanorings or nanodisks. New phase transitions have been evidenced in these nanostructures thanks to this highly sensitive device.

3.1.1 Experimental set-up for attoJoule calorimetry

The sensor is composed of a silicon membrane roughly ten microns thick. On this membrane two transducers are deposited by microfabrication technique: the copper heater and the niobium nitride thermometer [9]. The empty space between the two transducers is used to deposit nanosystems by ebeam lithography. As shown in the Fig. 1, the free space totalized 4 mm$^2$ which permits the deposition of a very high number of identical nanostructures totalizing 100 nanograms, as we will see below, at the limit of the detection of the device. As explain in the first section, one of the first characterizations of the sensor is to define the adiabatic plateau in order to know in which frequency range the sensor can be considered as quasi-adiabatic. One such typical adiabatic plateau is shown in the Fig. 2. As the maximum frequency can be easily seen, this plot is far from being academic. The bell-shape of the response indicates that the system is thermally isolated from the heat bath above 180 Hz. At high frequency, it can be noticed that there is an intrinsic limitation due to internal diffusion of heat. This may explain the decrease of the response above 180 Hz. In our case the working frequency for the heat capacity measurement has been chosen between 180 and 220 Hz. The resolution of the experiment is estimated to be $\Delta C/C = 5 \times 10^{-5}$ in the best situation. This allows the measurement of the heat capacity variations associated to nanosystems involving small mass.

This resolution is reached when the noise is only limited by the noise of the thermometer which should be close to the Johnson noise. Because around 1K the heat capacity of the total sensor is around 100 pJ/K, a heat capacity variation of 5 to 10 femtoJoule/K can be measured. The energy involved in this measurement corresponds to the amplitude of the temperature oscillation times the heat capacity resolution which gives energy sensitivity of the order of few attojoule ($10^{-18}$ J) [10].

This extreme sensitivity has been reached thanks to the ac measurement based on a lock-in amplifier technique and very low noise preamplifier. This method can be applied down to very low temperatures with an intrinsic limitation related to the continuous temperature $T_{dc}$. This dc temperature prevents the experimenter to measure at the minimum temperature provided by the refrigerator like a dilution or $^3$He cryostat, this may be a strong limitation for this technique if one wants to make ultralow temperature measurement of heat capacity (below 10 mK) [60].

3.1.2 Measurement of nanosystems: superconducting second order phase transition at the nanoscale.

This method has been applied for the measurement of superconducting nanorings and nanodisks in order to study the impact of the reduced size on the heat capacity jump at the second order phase transition [11,12]. It is indeed expected that when sizes are reduced specific behaviour may happen in the heat capacity signature occurring at a phase change [13,14]. The phase diagram may be modified but also the ordering temperature is evolving as compared to the bulk one. Here we are more interested in the change of the amplitude of the jump when an external parameter like the magnetic field is applied. We show that depending on the size of the nano-objects the heat capacity jump does not depend only on the total mass of the sample but also on its topology.

Two types of systems have been studied having different topologies: doubly connected superconducting rings and simply connected superconducting disks. The first experiment was performed on 2.5 millions superconducting aluminium rings of 1 μm in diameter, 30 nm thick and having arm width of 180 nm for an equivalent mass of about 100 nanograms [11]. The heat capacity versus temperature of the rings has been measured for different applied magnetic fields. The magnetic field is applied normal to the rings. The critical temperature is extracted from the middle of the transition, at zero magnetic field $T_c$=1.46 K. A clear second order phase transition has been detected from which the heat capacity jump $\Delta C_\phi$ is extracted. The interesting physical parameter is the magnetic flux $\phi=BS$, the flux included inside the superconducting ring. It is known since the 60's that the magnetic flux is quantized in such a loop to a integer number of magnetic flux quantum given by $\phi_0=h/2e$ where h is the Planck constant and e the charge of the electron. The variation of the heat capacity jump has been plotted as a function of magnetic field in the Fig. 3. We show here that the quantification of the magnetic flux has a strong influence on the heat capacity jump occurring at the second order phase transition between the superconducting state and the normal state. The magnetic field value corresponding to one quantum of flux in these rings is around 3.2 mT. This is approximately the periodicity of the oscillations observed in the Fig. 3. The ring shape of the nanosystems has a major influence on the heat capacity signature of the second order phase transition. This is one of the first illustrations of the topology effect on the thermodynamic signature of a phase transition. However due to the very small mass of the sample the error bars are still big. We decided then to measure a sample having a larger mass. Henceforth, the second experiment has been carried out on an assembly of $4 \times 10^5$ superconducting disks of 2 μm in diameter and 160 nm in thickness involving a total mass of aluminium of 600 nanograms, which is much bigger than in the previous experiment [12]. The critical temperature for such massive nano-objects is around 1.26K closer to the critical temperature of the bulk materials (1.2K). As in the case of superconducting rings, the heat capacity jump has been measured versus the temperature for different value of the magnetic field with a higher accuracy than for the rings. It is due to the larger mass of the sample and hence the jumps have larger amplitude; they have been plotted as a function of magnetic field H in the Fig. 4. As the error bars are smaller than the size of the points then the evidence of a modulation of the heat capacity jump with an external parameter is quite convincing. Here we cannot related the periodicity of the variation to a specific value of the magnetic flux because, due to the simple connected geometry of the disk, the magnetic flux quantum is not uniquely defined and then we do not control the exact repartition of the vortices in the surface of the disk. However, the variation of the heat capacity jump versus the magnetic field is really significant; variations up to 30% have been observed confirming what has been evidenced in the previous experiment for the superconducting rings.

To summarize, for these small systems it has been demonstrated that the topology and the reduced size do matter significantly when dealing with the heat capacity jump at the superconducting second order phase transition. The heat capacity jump can be modulated or changed in any second order phase transition by applying an external parameter. For instance this may be the case for ferroelectric; the application of an external electric field could be used to tune the thermodynamic signature of a second order ferroelectric phase transition. The lesson to be taken out is that at the nanoscale the heat capacity jump, signature of a second order phase transition is not only given by the mass of the materials involved but also by the geometry and the size of the system considered especially when it is scaled down to the nanometer. An other calorimetric technique exists which generally at low temperature is based on relaxation method [15]. However, it is not yet clear if this method has sufficient sensitivity to allow the measurement of nanosized sample down to very low temperature [16].

*3.2 Polymer membrane based ac calorimetry*

In the intermediate range of temperature of 20 to 300K, not so many sensors exist allowing highly sensitive calorimetric measurements of nanoscaled samples [17,18]. Here we present a sensor dedicated to the measurement of magnetic thin films under high magnetic field based on polymer membrane (polyparaxylylene; Parylene C) [19]. This material has the major advantage of having a very small heat capacity as well as being quite inert in this temperature range. We show that high sensitivity may be reached which opens up large applications for this sensor.

3.2.1   Sensor fabrication

In the range of temperatures between 20 and 300 K, a polymeric membrane-based ac calorimeter has been developed to measure magnetic phenomena in small samples of around 1μg mass, like: microcrystals , thin films or nanoclusters, see Fig. 5. The calorimetric device is built-up on a copper substrate of 20 mm diameter and 300 μm thickness that acts as a heat sink. An initial buffer layer of 5 μm of polyparaxylylene (Parylene C) is deposited on the copper substrate after a polishing step. On top of this insulating layer, the calorimetric cell is defined as a stacking three layer sandwich. In first level, a 100 nm thin film layer of niobium nitride alloy with a low nitrogen concentration ($NbN_{1.1}$) is sputtered and lithographed to define the heater element. After that process, at a second level a 500 nm thin layer of polyparaxylylene is deposited. This layer is the support for the thin film transducers and it withstands the mechanical stress of the structured cell. Electrically insulated from the beneath element by the last polyparaxylylene layer, in a third level the thermometer is defined by a standard lift-off process, using Au contacts to prevent contact noise between the $NbN_{1.1}$ used as current leads and the highly nitrogen doped $NbN_{1.7}$ that is used as a high sensitivity thermometer.

At such high concentration of nitrogen this material presents a metal-to-insulator transition at relatively high temperature, and by tuning the nitrogen content it is possible to obtain the maximum temperature coefficient of resistance (TCR) in the required temperature range [20]. Next, the membrane is freed, removing a window in the copper substrate with iron perchloride acid. The 5 µm thick polyparaxylylene and the non protected by the NbN structure thin polyparaxylylene is removed by oxygen plasma, leaving the final thin film three layers that defines the calorimetric cell, with a sensing area of 1 mm$^2$, suspended by arms of 0.1x0.1x1mm$^3$. A schematic view of the finished sensor is shown in the Fig. 5 where the three layer systems can be seen in the top panel Fig. 5a, the top view show the geometry of the measurement cell itself in Fig. 5b, and finally a global photograph gives an idea of the cell and the electrical contacts used for the ac measurement of the thermal property of the calorimetric device.

3.2.2 Sensor characterization

This geometrical layout combined with the selection of poor thermal conductive materials (like NbNx and polyparaxylylene) permits to achieve adiabatic conditions at very low frequencies (around few Hz) using the ac method for calorimetry [19]. Fig. 6 shows the dynamic response ($\omega \cdot \delta T_{AC}$) of the calorimeter versus frequencies. At low frequencies the signal is essentially driven by the conductive component of heat flowing from the calorimetric cell to the copper frame that acts as a heat sink. Increasing the frequency, the dynamic response ($\omega \cdot \delta T_{AC}$) is enhanced, since the adiabatic conditions for the ac components are improved. However at high frequencies it starts to degrade when the limit of in–plane thermal diffusion is reached. The maximum defines the adiabatic working frequency $f_{AD}$, in a range of frequencies that permit in the future to perform magnetocaloric experiments by modulating the magnetic field at low frequency. For the measurement of heat capacity, the sensor is loaded in a He cryostat that permits to fix the baseline temperature in the range between 4 and 300 K, and also assures a cryogenic vacuum in the surrounding of the calorimetric cell. Typically, the heater is excited at the optimum frequency with a custom built low noise ac current source to promote amplitudes of temperature oscillation between 1-50 mK. Using a lock-in amplifier to measure the second harmonic and due to the high sensitive thermometry, the temperature signals can be measured with sensitivity better than $10^{-3}$, which means we are able to detect few µK variations in the signal amplitude.

In Fig. 7, the heat capacity is shown as function of temperature for an empty calorimetric cell, with a point taken every 50mK in the range between 160 and 220 K. As it can be observed in the inset of the Figure 7, the sensitivity permits to resolve features in the nJ/K range, which permits to detect energy variation of 10 picojoule considering that the amplitude of the excitation temperature is of the order of 10 mK. This sensor will be used in a near future for the measurement of nanomagnetic materials as well as magnetic thin films in order to localize precisely the presence of a second order phase transition in the case of ferromagnetic or antiferromagnetic nanoclusters or very thin films.

**4. Ambient temperature nanocalorimetry**

*4.1 Polymers*

Polymers are known to be complex systems because of the high numbers of atoms composing such macromolecules. The physical and chemical behaviour of such systems are still not understood [21]. Polymers are particularly interesting because their physical properties may be a link between condensed matter systems and biological macromolecules (part3, part 4.2). In this part, we study the thermodynamic behaviour of a well-known polymer, the polytetrafluoroethylene (PTFE) which undergoes two phase transformations at ambient temperature. As the intrinsic kinetics of the two phase transitions are of very different nature, this polymer becomes a physical model helping our understanding of the dynamic aspect of the calorimetric method.

4.1.1 Nanocalorimetric device.

A specific micro-fabricated nanocalorimetric device has been designed for the measurement of small film of solid polymer. Since all the experiment is already described in details in the references [6] and [22], we recall here only the principal features of this sensor and this experiment. On the drawing of Fig. 8, the sample is held between two stainless steel thin membranes (12.5 μm thick). On the sides of the metallic membranes not in contact with the sample, thin polyimide films (5 μm thick) are spin coated. A thermometer is micro-patterned on the bottom film, and a heater on the top film. Each stainless steel membrane is glued on a hollow copper cylinder which serves as a constant temperature bath for calorimetric measurements. The top and bottom closures of the copper cylinders are situated few millimeters back of the metallic membranes. The volume between the copper closures and the metallic membrane is connected through a small pipe to a tank of 2 litres filled with gaseous nitrogen under pressure of 1 or 2 bars. This tank which is regulated in temperature is outside the calorimeter. The two copper cylinders are tightly clamped on a massive copper piece. This piece is regulated in temperature by means of a thermometer (high precision Pt100 resistor) and a heater. The whole is thermally linked to a Peltier element which is the cold source of the experiment. The thermometer and the heater are included in a servo-system which allows the temperature of the entire cell to follow the temperature ramps, or to be regulated at a constant temperature. The precision is about 0.1 K and the noise is about $10^{-4}$ K. All these elements give a heat capacity resolution $\Delta C/C = 10^{-5}$. With the absolute value of the total heat capacity of about 20 mJ/K, this calorimeter allows the detection of thermal events as low as 200 nJ/K. With typical amplitude of temperature oscillation of 0.2 K, this corresponds to an energy detection of about 40 nJ. For a sample thickness of 50 μm the heat capacity of the empty sensor (addenda) is equivalent to those of the sample ($C_{add} \sim C_{sample} \sim 10$ mJ/K). For a sample thickness of 100 μm used for low frequencies measurements, the heat capacity of the sample is twice that of the addenda.

4.1.2 PTFE : a model for out of equilibrium phase transitions in polymer.

The PTFE is known for undergoing two well distinct phase transitions around the ambient temperature [23,24]. The PTFE goes from the phase II to the phase IV around 293 K, and from the phase IV to the phase I around 303 K. The "first phase transition" from phase II to IV is mostly due to the twist of the polymer chains around their symmetry axis. This phase transition is mostly regarded as a first order phase transition because of the large hysteresis between the up and down scans. The "second phase transition" from phase IV to I is mostly due to molecular conformational transition (condis transition) with disorder all along the chains. It looks like a second order phase transition and the up and down scans seem perfectly reversible. However, a closer look on the curves reveals a small hysteresis between the up and down scans. This suggests that the two phase transitions in PTFE are macromolecular structural transitions with different kinetics. In the Fig. 9 two up scans of the heat capacity as a function of the temperature are provided, one obtained by differential scanning calorimetry (DSC) (microDSCIII from SETARAM Instrumentation) at scanning rate of 0.2K/min and the other obtained with the ac-calorimeter at a frequency of 0.2 Hz with a underlying scanning rate of 0.25K/min. Only few percent of the first peak is measured by ac-calorimetry, revealing the out of equilibrium nature of this temperature induced phase transition in PTFE. The second peak does not change with both methods. This kinetic effect has already been observed by means of ac-calorimetry on the PTFE [25] and frequency effects have already been reported at low frequencies with TMDSC [26, 27]. For the presented ac-calorimetric device, we have worked out a thermal model for the sample and the sample holder which enable us to extract data in a larger frequency range (0.020Hz-40Hz) than the frequency range given by the inequalities (4). The raw data issued from the measurements of amplitude and phase of the oscillating temperature at different frequencies, as well as the description of the thermal model used in order to transform these data in real and imaginary parts for the heat capacity, are already described in the reference [6]. Here, we are just presenting the complex heat capacity components C'(T) and C"(T) for the PTFE at three distinct frequencies (0.04Hz, 0.4Hz, 4Hz) on Fig. 10.

We emphasize in this paper the out of equilibrium behaviour of the polymer probed with ac-calorimetry. We observe that even for the "second phase transition" of the PTFE an imaginary heat capacity component can be extracted from the measurements. This tends to prove that this second phase transition is of the same type that the first one but with smaller relaxation time. Let us remember that the imaginary part of the measured complex heat capacity reveals directly the entropy produced during one period of the temperature cycle [5].

This mean production of entropy is due to internal friction or heat dissipation inside the system while it cannot follow instantaneously the temperature oscillation. The relation between the imaginary part of the frequency dependent heat capacity and the mean entropy production has been investigated in detail for different experimental situations in the reference [28] for ac-calorimetry, and in the reference [2] for TMDSC. It has also recently be shown that the entropy production generated in a slow relaxing system submitted to a temperature perturbation can be regarded as the irreversible exchange of heat between two sub-parts constituting the system [29 and references therein]. The dynamic character of the PTFE, mostly for the "first phase transition" has already been investigated by TMDSC [26]. Our results, albeit they are at higher frequencies, are in good agreements with TMDSC results. One interesting point in the PTFE is the possibility to proceed to thermal spectroscopic experiments knowing that the frequency effect happening on the "first phase transition" is not due to experimental pitfalls, because the "second phase transition" is rather frequency independent. Finally, our data cannot be fitted by a simple Debye relaxation equation such as (5), which means there is not a single relaxation time involved in the "first phase transition" in the PTFE. Nevertheless, when a classically modified relaxation equation is used for the complex heat capacity:

$$C^* = C_\infty + \frac{C_0 - C_\infty}{\left[1 + i(\omega\tau)^{1-\alpha}\right]^\beta} \tag{6}$$

the data are very well fitted with $1-\alpha = 0.55$ and $\beta = 1$.

### 4.1.3. Conclusion and perspectives

This first study of the dynamic behaviour of phase transitions in a polymer by means of ac-calorimetry has demonstrated that it is possible to access directly to thermodynamic quantities such as the heat capacity or the entropy when the system is so rapidly constrained that it is brought out of thermodynamic equilibrium. As the ac-calorimetry is a stationary method of measurement when the frequency is changed, then the time scale of observation is changed and the experimenter accesses to different thermodynamic states of the sample around a mean dc temperature. It is a true spectroscopic technique. The underlying slow dc temperature ramp allows us just to change the surrounding conditions such as the temperature of the reservoir. In classical DSC the situation is totally different because in this case the scanning ramp is directly implied in the measured thermodynamic quantity. In a close future, in order to work at lower frequencies still remaining adiabatic we plan to use a new nano-calorimetric sensor where the stainless-steel thin membranes will be replaced by polyimide ones. Such a sensor is shown on the Fig. 11. Consequently the electronic conditioning has to be modified, particularly a very low noise, low drift and low frequency pre-amplifier has to be developed. At such frequencies systematic comparison between ac-calorimetry and TMDSC on a same sample will provide new information on slow kinetics of phase transitions in polymers.

*4.2 Biological objects*

4.2.1 Protein folding/unfolding

4.2.1.a Introduction: scientific perspectives

The understanding of the folding/unfolding of a protein is an old but still unresolved question. How a macromolecular system composed of few thousand of atoms can spontaneously, under physiological conditions, folds into a well-defined three-dimensional structure under a time scale compatible with biological life ? What is the driving force of protein folding processes ? The answers to these questions are of importance not only from a physical but also from a biological and biochemical points of view. In medicine, the misfolding of specific proteins results in the development of several disorders such as Alzheimer's and Parkinson's diseases. For physicists the large range of time scales involved during the folding pathway makes these systems quite interesting. Rapid fluctuations of protein arise within the ps to ms range. Global folding involves different steps occurring on timescales of the order of ms to minutes. These different time scales can be understood by using the energy landscape approach [30]. Fluctuations occur around minima in the energy landscape whereas global configurational steps occur between these minima. These jumps between different minima are kinetically driven. It is already well-known that classical calorimetric experiments (DSC) coupled with spectroscopic experiments (UV absorption, circular dichroïsm, etc) are powerful tools in order to study such phenomena. The dynamic aspect of the ac-calorimetry allows us to study or to scrutinize different underlying events during the thermal unfolding of the protein. Now, on a pure experimental point of view, general calorimetric techniques such as DSC are not enough sensitive to be competitive in terms of quantity of proteins used for measurements as compared with spectroscopic techniques. To give an order of magnitude, the best available DSC apparatus need 1mg/ml concentration of protein in buffer to obtain reliable results while spectroscopic apparatus need ten times less protein concentration. Moreover, calorimetry generally needs high cell volume to have enough material to obtain thermal signatures. The best available DSC needs sample volume around 300 μl, but usually cell-volumes are close to 1 mL. An ac-nano-calorimeter might avoid these two problems, because firstly the ac-calorimetry is the most sensitive calorimetric method of measurement (although now high speed DSC are competitive [14, 31, 32]), and secondly micro-fabrication technologies allow reduction of addenda and thus a reduction of sample volume is possible. Increase in sensitivity and reduction of sample mass contribute to the same result, i.e, a decrease of quantity of biological molecules for calorimetric measurement which is of great interest for biochemists.

4.2.1.b Nanocalorimetric device for liquids.

A specific ac-calorimetry cell has been designed by means of micro-fabrication technologies. The experiment is described in the reference [33]. Here we outline briefly the features of the experimental device. The measurement cell shown in the photography of Fig. 12 is composed of two half-cells. Each half-cell is made of one thin polyimide membrane (Kapton® polyimide film, type HN from Dupont company) of 12.5 µm thick pasted on a copper massive piece serving as thermal bath. The thermal insulation between the polyimide membrane and the copper holder is obtained via few thousand of micro-posts such as seen in Fig. 13. Heat flows principally across the micro-posts with a total value of thermal conductance of 20 mW/K. The lateral heat flow is stopped by means of a thermal open circuit constituted by a thin trench etched in each membrane such as shown in the Fig. 14. A platinum thermometer and a copper-nickel heater are deposited, micro-photolithographied and patterned on each half cell. Generally one half-cell supports the heater whereas the second supports the thermometer, but sometimes a heater and a thermometer are both present on one half-cell and interdigited like shown in Fig. 14. A micrometer thick polymer film is spin coated on these metallic films in order to insure electrical insulation between such sensitive elements and the liquid sample. For the measurement, the small amount of liquid is enclosed between two half-cells sealed by a thin polymer gasket of 50 µm thick (Teflon® FEP film, type A from Dupont company). The thickness of this ring determines the thickness of the sample. Usually a thickness of 50 µm is chosen, but sometimes a thickness of 100 µm can be used when low frequency measurements are planned. Knowing that the internal diameter of the gasket delimits a surface of 1 cm$^2$, the volume of the chamber for a thickness of 50 µm is only of 5 µl. With about 20 mW/K for the thermal conductance and 30 mJ/K for the measured heat capacity (about 20 mJ/K for the sample and about 10 mJ/K for the addenda), the condition of quasiadiabaticity is fulfilled for a frequency around 3 Hz. At this frequency due to the small thermal diffusivity (about $10^{-3}$ cm$^2$/s) and the small thickness of the liquid the whole sample volume oscillates in phase. The frequency response of such a sensor (filled with water) is shown in Fig.15.

Once the liquid sample is enclosed hermetically in the cell, the whole cell is loaded in a calorimeter under vacuum for preventing parasitic heat exchanges and favor temperature control. By means of a small lateral hole, the air between the micro-posts is removed preventing additional heat transfers. The temperature of the cell is controlled by a thermoelectric cooler and adequate PID regulation system stabilizes the temperature better than 1 mK in static or dynamic regimes. The oscillating voltage is amplified with a low noise pre-amplifier (1 nV/$\sqrt{Hz}$ in the frequency range of 0.1 to 100 Hz) which enable us to work at the level of the white noise of our thermometer (about 1 kΩ at room temperature). We then obtain a resolution in heat capacity of $\Delta C/C = 10^{-5}$. This yields to a limit of detection of $\Delta C \sim$ 300 nJ/K. For an amplitude of temperature oscillations around 0.5 K, this gives a limit of detection in energy around 150 nJ. The noise performances of the nano-calorimeter are detailed in the reference [33].

4.2.1.c Thermal denaturation of the lysozyme

It is known that thermal signatures during thermal unfolding of proteins are very weak [34, 35]. Moreover, it is also known that in the unfolded state proteins tend to aggregate at high temperature. Here we emphasize three points. Firstly, the ac-calorimeter reveals thermal denaturation of proteins. Secondly, the aggregation effect is also seen with the ac-calorimeter like by DSC. Thirdly, and it is the most surprising feature, at a frequency of 3 Hz the process of folding/unfolding is so slow that we do not have access to the enthalpy peak of the thermal denaturation of the lysozyme as classically obtained by DSC. At such a frequency we have only access to the specific heat associated with the fast degrees of freedom as presented in section 2.

The heat capacity of two solutions of lysozyme (lysozyme from hen egg white, lyophilized powder from Fluka company, 107648 U/mg; Mr~14 600) diluted in adequate aqueous buffer solution (pH=3 and pH=4.5) is plotted versus the temperature in Fig. 16. The solution at pH = 4.5 contains 25 µg of proteins whereas the solution at pH = 3 contains 40 times more of proteins (1 mg of proteins). The raw data do not reveal impressive effects, although for the high concentrated solution a slight discontinuity in the alignment is visible. Owing to the high sensitivity of the measurement, it is however possible to mathematically derivate the curves in order to reveal the thermal events. This is depicted in Fig. 17. Following these first tries, the intermediate conclusions are threefold:

-the ac-calorimeter gives access to the thermal unfolding of the lysozyme for the two concentrations used.

-the effect of concentration is clearly visible.

-the pH effect is also detected because we observe that for pH = 4.5 the maximum of the peak in the derivative occurs around 75 °C, while for pH = 3 it is mostly centred around 64

°C. This reveals a loss of thermal stability of the protein due to acidic buffer solution. However, we can observe that even for such high quantity of proteins as 1 mg in the cell, the protein thermal denaturation is difficult to observe. We have also made successive up scans of the same solution of lysozyme (pH = 3) at high concentration. The Fig. 18 shows the mathematical derivatives of the C(T) curves obtained. The derivative of the C(T) buffer solution is also shown for comparison. An irreversible effect attributed to the aggregation of the lysozyme is clearly seen as a function of the successive scans. A decrease of the peak is seen at the same time than a spreading over a broader temperature range. In the Fig. 19 the differential signal of a sensitive DSC (Nano-DSC-III CSC-TA Instrument) reveals also the aggregation effect during thermal unfolding of the protein after successive up scans.

The decrease effect depicted in the two previous figures looks like identical. Nevertheless, curves in the Fig. 18 are derivatives of the heat capacity whereas those in the Fig. 19 are directly proportional to the heat capacity of the proteins. At a frequency of 3 Hz, with ac-calorimetry we do not have access to the usual enthalpy peak observed by DSC. Only the heat capacity jump during the thermal unfolding of the proteins, such as represented by the baseline of the Fig. 19, is accessible. Thus, only the heat capacity of the vibrational modes of the protein (and interactions with solvent) is recorded when the lysozyme passes from its folded state to its unfold one ($C_\infty$ term of the Eq. (5)). We are now able to measure the heat capacity jump $\Delta C_p$, which during protein folding/unfolding determines the curvature of the stability curve $\Delta G(T)$. Our next experimental step is to have access to the dynamic of the thermal denaturation by increasing our working frequency range.

4.2.1.d Conclusion and perspectives

Dynamic effect during thermal unfolding of the lysozyme has already been observed by other groups with high sensitive ac-calorimeter and TMDSC [36-38]. From reference [38] at a frequency of 3.3 mHz, all the enthalpy component has been measured. From reference [36], at a frequency of 50 mHz only 60% of the enthalpy component has been observed. From our measurement, at a frequency of 3 Hz, the enthalpy component is not detectable. It has to be noticed that this qualitative consideration does not take into account the difference of physiological conditions (buffer, pH, concentrations, etc...) used in the different experiments and also the different calorimetric methods of measurement. Anyway, up to now no spectroscopic calorimetric measurement has been performed on such systems. Due to the aggregation process, a new sample has indeed to be filled into the calorimetric cell for each experiment. It is particularly difficult to have perfectly identical sample. Also, due to this aggregation process, the cell has to be especially well-cleaned out between two simple up scans. A process of easy cell-filling and cleaning would be a great improvement.

4.2.2 Living cells

4.2.2.a Introduction: scientific perspectives

Living matter is a very intriguing state. Unlike inert matter, living systems are open systems continuously evolving due to the exchange of matter, energy and entropy with the surrounding medium. The extremely high degree of organization of living object already studied by D'Arcy Thomson [39] is surprising and seems at first sight to contradict the second law of thermodynamics [40]. In 1970 based on a calorimetric analysis of the catabolism processes of bacteria (the processes that degrade highly organized form of nutrient to form

ATP), the outflow of entropy from the living systems was deduced [41]. From these early studies it was concluded that most of the energy necessary for cell growth is that required for correct placement of the molecules within cells.

Bacterial growth can be basically described with the linear non-equilibrium thermodynamics formalism [42,43]. Within this framework the entropy change for an open system is:

$$dS = d_e S + d_i S = \frac{dQ_e}{T} + (d_e S)_{matter} + d_i S \qquad (7)$$

where the variation of entropy due to the exchange of matter $(d_e S)_{matter}$ takes into account the flow of metabolites and the synthesis of low entropy highly organized biomass. To abide the second law of thermodynamics if the living system can be considered as stationary then on one hand the entropy creation of the system should be minimum on the other hand the first two terms of the right hand side of the equation (7) should compensate. The out of equilibrium state in which living systems are is an essential property that leads through the continuous chemical degradation of nutrient toward the establishment of a very organized state, with a driving force [44] equal to the entropy production $d_i S$ that is accessible with calorimetric experiments. The analysis of individual metabolic pathway is a very interesting fundamental perspective of such experiments [44]. Another more applied very interesting perspective would be its application to the pharmaceutical modulation of the metabolism of mammalian cells for the identification of relevant drugs [45] or even potential infections [46].

Progress in microfabrication are now allowing the design of nanocalorimeters for the monitoring of the metabolism and specific activity of small populations of cells [47,48] and even single cells [49] are also being designed for the measurements. All the nanocalorimeters developed for that purpose used thermopile as a thermal sensor. The best achievement being the one of Lerchner *et al.* [50] with a signal to noise ratio of less than 10 nW and a volume of about 6 μL.

We propose to use the polymer membrane technologies and the high sensitivity and stability of platinum thermometry to build up a few cells calorimeter able to monitor the change in metabolism in stationary and non stationary mammalian cell cultures. We present the design of an isothermal power compensation differential calorimeter able to monitor thermal signature of single cell events. For maximum sensitivity we require an excellent coupling of living cells with the thermometer. For that purpose we developed a nonspecific immobilization technique for cells to specifically adhere onto the sensor. We present the immobilization procedure first. Then the basic concept associated with the sensor design will be explained. The perspectives and calibration test to be done will be presented at last.

4.2.2.b Non specific protein immobilization procedure for cell adhesion on physical sensors

Cell adhesion is a fundamental process of cell biology. It is involved in cell differentiation during the embryo development [51] and many pathologies particularly those associated with cancer metastasis [52]. Living cells are complex systems regulated biochemically. The specific interaction of the proteic receptors present on the cell membrane with the proteins forming the extracellular matrix triggers the biochemical signalling that regulate the cell fate and function [53].

The methods for proteins immobilization are described in [54]. In short, cleaned glass cover slips (22×22 mm$^2$) were dipped for 1 h in a solution of octadecyltrichlorosilane and hexane 0.2% v:v. The SiCl3 group of the alkylsilane reacts with the silanol groups present on the glass surface and optical measurements of water contact angles between 100° and 110° indicate the formation of robust multilayers of long hydrophobic carbon chains [55]. Positive photoresist was spin-coated and cured according to the manufacturer's protocol to form a uniform UV-sensitive film 0.5 μm in thickness. The cover slips coated with photoresist were then irradiated with UV light at 436 nm and 15 mJ/cm$^2$ through a chromium mask, and the irradiated pattern was revealed. The glass cover slips with patterned photoresist were incubated for 1 h at 37 °C in a solution of the Fluorecein isothiocyanate (FITC)-labeled cellular interaction fragment of fibronectin made of type III domains 7 to 10 (FN7–10) [56] at the concentration ranging from 5 to 15 μg/ml in PBS. Nonspecific adsorption on glass or silicon of fibronectin is approximately self-limited to a monolayer [57] and the hydrophobicity of the substrate increases the protein packing and consequently the layer homogeneity [58]. The substrates were then rinsed in PBS and subsequently in pure ethanol in an ultrasonic water bath to dissolve the photoresist resin. Finally, on the complementary pattern revealed after resist dissolution by ethanol, antiadhesive tri-block polymer Pluronic F127™ (Sigma-Aldrich 4% in water, 1 h 30 min at 37 °C) were adsorbed. A final rinsing in PBS is the final step of the pattern microfabrication of the cover glasses now ready for use.

Living cells (NIH 3T3 Fibroblasts) were plated onto the functionalized substrates and let for spreading for 4 hours. Longer term cultures were possible for 3 days before the protein secreted by the adherent cells would blur the micropattern. The Fig. 20 shows typical image of a fixed cell on a non patterned substrate (Fig. 20 a) and on a patterned substrate (Fig. 20b). The micropattern is a very useful tool for the control of the geometry and positioning of cells onto a surface. The Fig. 21 shows an array of cells with almost optimum filling. These tools are required if quantitative data are wanted. Indeed as already noted the internal organization of cells is most likely to be responsible for most of the energy dissipation in growth [41].

4.2.2.c Nanocalorimetric device.

The nano calorimetric device will incorporate two sensitive zones with sputtered thick film platinum resistors inserted in a Wheatstone bridge and connected to a high sensitivity and stability electronics presently under test. The Fig. 22 represents a schematic of the experimental setup. The isothermal calorimeter will operate in differential mode and in the quasistatic mode $\Delta P = P_s-P_{ref} = K\Delta T$ limit since the events to be measured take place on timescales much longer than the thermal relaxation time constant of the calorimeter. Under such specific conditions we benefit also of the inherent low pass thermal filter of the nanocalorimeter. Thus the sensitivity and stability on the temperature measurement fixes the minimum power that can be detected. The temperature is measured through a resistive thermometer and the limiting noise is the resistor Johnson noise:

$$\delta V = \sqrt{4k_B TR\Delta f}$$

The corresponding noise in temperature is: $\delta T = \dfrac{\delta V}{\alpha V}$ ($3\times 10^{-7}$ K/$\sqrt{Hz}$) where $\alpha=(1/R)(dR/dT)$ = $3.9\times 10^{-3}$ K$^{-1}$ for platinum. We then obtain a differential temperature detection better than 1 µK/$\sqrt{Hz}$ at ambient temperature if the noise of the thermometers is closed to the white noise (1 nV/$\sqrt{Hz}$). The power noise is then $\delta P = K\delta T$. For the value of the parameters of the sensor and the bridge that we designed this leads to the minimal power that can be detected of the order of $\delta P \approx 3$ nW/$\sqrt{Hz}$. According to Kemp [59] the metabolism of living cells range between 1-100 pW. Thus to detect 3 nW we would need to build arrays for 3000 to 30 cells. Since events to be detected are very slow, time averaging will allow us to reduce again the minimum detectable power.

4.2.2.d perspectives

The building of the presented calorimeter is under way. Most of the effort has been concentrated on the placement on a controlled field of living cells. This is now mastered and the design and test of the calorimetric sensor and of the electronics is under progress. These are the major challenging issues to be undertaken in the near future. It will allow us to measure the metabolism of 30 to 3000 cells for fundamental purposes mostly related with out of equilibrium thermodynamics, diagnostic and pharmaceutics.

5. **Conclusions**

The use of the micro and nanofabrication facilities has opened up new area in sensitive nanocalorimetry. In this review on our current researches, different membrane based nanocalorimeters have been presented. They all have been designed for sensitive ac-calorimetric measurements. The underlying physics to be tackle determines the thermal design of the sensor. For low temperatures physics, very thin sensors have been built for the detection of very small heat capacity variations. At room temperature, stronger polyimide membranes are necessary to hermetically enclose a small amount of liquid without evaporation. For all these sensors the thermal insulation from the surrounding is crucial. We have discussed the possibility of measuring metabolism of single leaving cell. The first step consisting of controlling a particular biological feature of living cells (adherence control) and of appropriately positioning living objects on a sensor is now under control. The future challenge is to be able to reach the nanowatt sensitivity in thermal power detection over time-scales of more than few days.

**Acknowledgements**

First we would like to express our deep gratitude to J. Chaussy for initiating most of these subjects of research. The authors would also like to strongly thank the technical staff for their technological prowess and great know-how: E. André and P. Lachkar for sensors microfabrication (Pôle de Capteurs Thermométriques et Calorimétrie), J-L. Bret for low signals and low noise electronics conditioning, C. Guttin for data acquisition systems (with the members of the Pôle électronique), and P. Brosse-Maron for calorimeters embodiments and designs (with the members of the Pôle de Cryogénie). Finally, we acknowledge the different financial supports helping us to perform such researches: Région Rhônes Alpes, ANR (Sensocarb, Quantherm).

**Figure Captions**

Fig. 1 : 1a : Silicon membrane based sensor for application in nanocalorimetry below 10 K. On top the streamer is the copper heater, on the bottom the four contacts allows the four probe measurement of the thermometer with high resolution. 1b: Aluminum superconducting ring of 1μm in diameter. 1c: Aluminum superconducting disk of 2 μm in diameter.

Fig. 2 : Adiabatic plateau obtained at 0.8K. The temperature oscillation is of the order of 0.1 mK at 200 Hz, the current in the heater is of the order of 0.1 mA and the current used to measure the thermometer is of 20 nA.

Fig. 3 : Heat capacity jump variation under the applied magnetic field for aluminium superconducting rings (see Fig.1b).

Fig. 4 : Heat capacity jump variation under magnetic field for aluminium superconducting disks. The disks are 2 μm in diameter and 160 nm thick (see Fig.1c)

Fig. 5: a) Schematics of the calorimetric cell, showing the three layers structure. b) Frontal view photograph of the calorimetric cell. c) General picture of the calorimetric sensor, where in the center is possible to appreciate the suspended structure by arms made in polyparaxylylene.

Fig. 6: Adiabatic plateau of a calorimetric sensor made on polyparaxylylene.

Fig. 7: Heat capacity versus temperature measured in an empty calorimetric cell every 50 mK. The inset shows that Gaussian dispersion of $\Delta C_p/C_p$ in the measure has a FWHM smaller than $10^{-3}$, which also defines its sensitivity.

Fig. 8: Schematic description of the cell functioning and design. See text for details.

Fig. 9 : Comparison of the excess heat capacity of the PTFE obtained by DSC and ac-calorimetry (Fth = 0.2 Hz) for a scanning rate of 0.2 °C/min for DSC and 0.25°C/mn for ac-calorimetry.

Fig. 10: Real and imaginary parts of the frequency dependant complex heat capacity of the PTFE in function of the temperature at three well distinct frequencies 0.04 Hz, 0.4 Hz, 4Hz.

Fig. 11 : Half-cell over which the heater is lithographed on a thin transparent polyimide freely suspended membrane of 25 μm thick in Kapton. The four measurement lithographed wires are visible on this optical photography.

Fig. 12 : Complete cell for liquid measurement. One contains the heater and the other the thermometer. A transparent thin spin coated polyimide film, less than 1 μm thick, electrically insulates the sensitive elements from the liquid sample.

Fig. 13 : SEM photography of polymer micro-posts ensuring thermal adiabaticity. The posts are spaced of about 178 μm and their eight is of about 12 μm. The gold layer is visible on the top of the posts.

Fig. 14 : Optical photography of the thermal open circuit avoiding lateral heat loss. Few micro-posts in yellow (gold) are visible in the bottom of the trench. The width of the platinum wire crossing the trench is only of 30 μm.

Fig. 15 : Adiabatic plateau of the nanocalorimetric cell with deionized water inside.

Fig. 16 : Heat capacity in function of the temperature of two solutions of lysozyme: one at a concentration of 5 mg/ml in a buffer solution of pH = 4.5 and the other at a concentration of 200 mg/ml in a buffer solution of pH = 3.

Fig. 17 : Mathematical derivatives of the two curves depicted in Fig.16.

Fig. 18 : Mathematical derivative of the C(T) curves for different up scans on the same lysozyme sample at pH = 3. The buffer C(T) derivative obtained in a previous experiment is also shown for comparison.

Fig. 19 : Four consecutive up scans of a solution of lysozyme diluted in adequate buffer solution measured by DSC with a scan rate of 1°C/min. A mathematical baseline (sigmoïde curve) is also shown. This baseline is adapted only for the first scan. It reveals the heat capacity jump due to the change of vibrational modes during thermal unfolding.

Fig. 20 : a) A fixed cell on a non patterned substrate. b) A fixed cell on a patterned substrate. The immunostaining reveals the organisation of the actin cytoskeleton in red, the adhesiv macromolecular scaffolds in green and the extra cellular matrix in blue. White bar is ten microns.

Fig. 21 : Array of cells with almost optimum filling. Such an array is used to put a large amount of living object in the same configuration according to external parameters.

Fig. 22 : Schematic of the isothermal calorimeter.

.

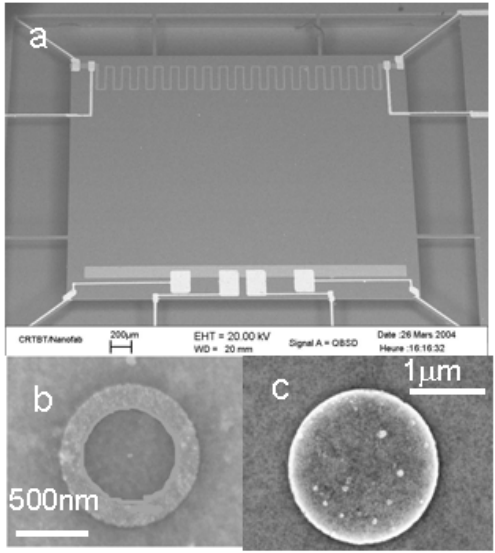

Fig.1 Thermochimica Acta. J.L. Garden

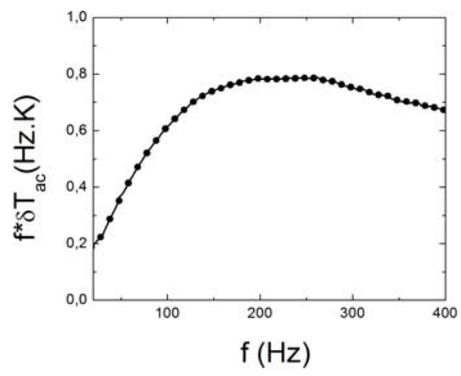

Fig.2: Thermochimica Acta. J.L. Garden

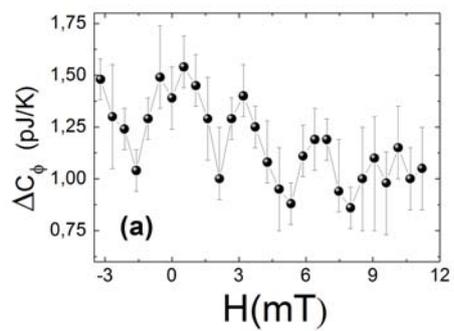

Fig.3: Thermochimica Acta. J.L. Garden.

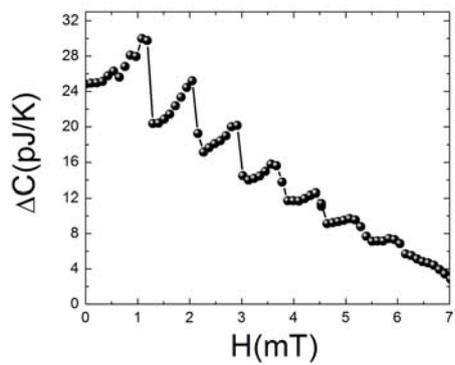

Fig.4: Thermochimica Acta. J.L. Garden

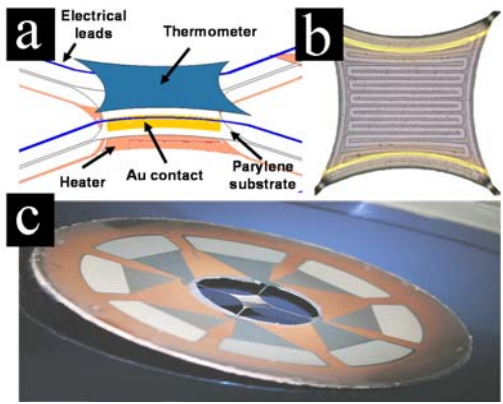

Fig.5: Thermochimica Acta J.L. Garden

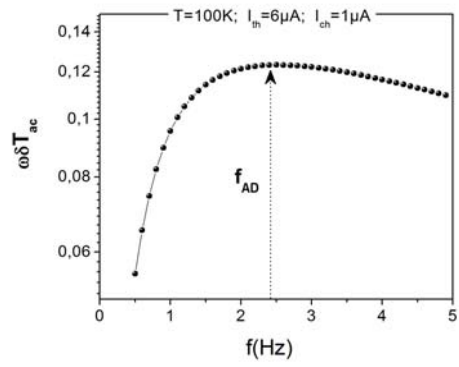

Fig.6: Thermochimica Acta J.L. Garden

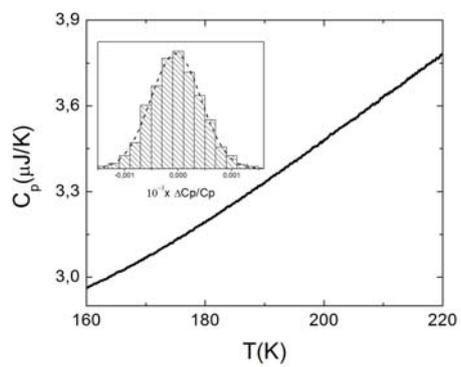

Fig.7: Thermochimica Acta J.L. Garden

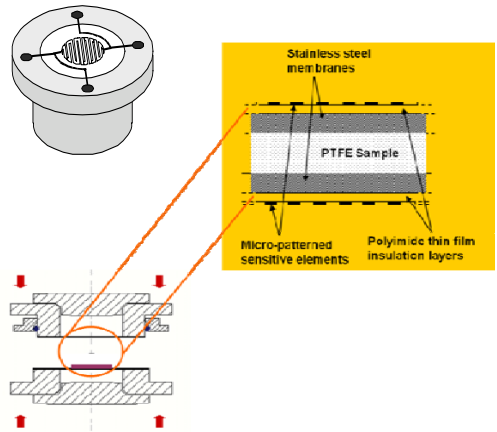

Fig. 8: Thermochimica Acta J.L. Garden

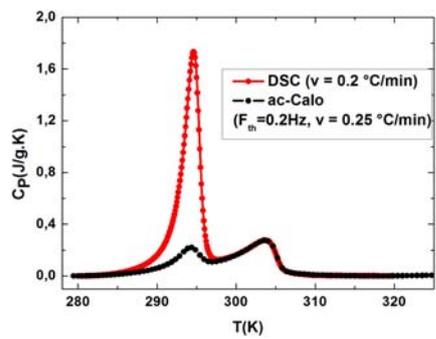

Fig.9: Thermochimica Acta J.L. Garden

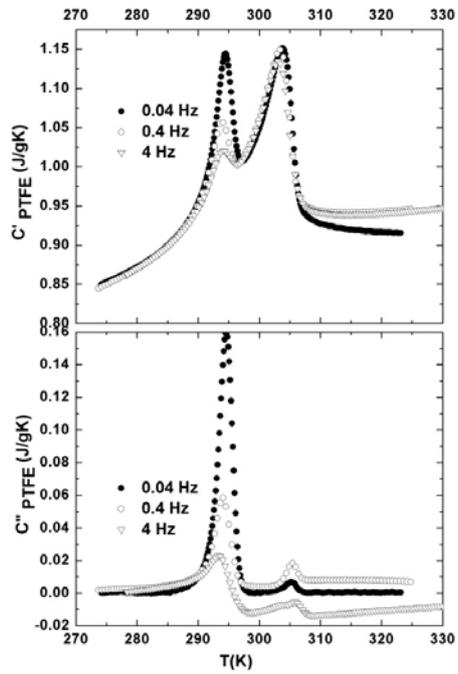

Fig. 10: Thermochimica Acta J.L.Garden

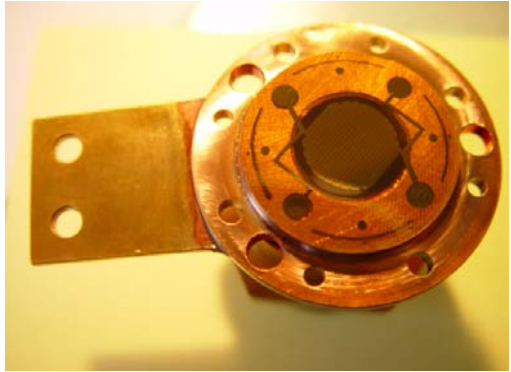

Fig. 11: Thermochimica Acta J.L. Garden

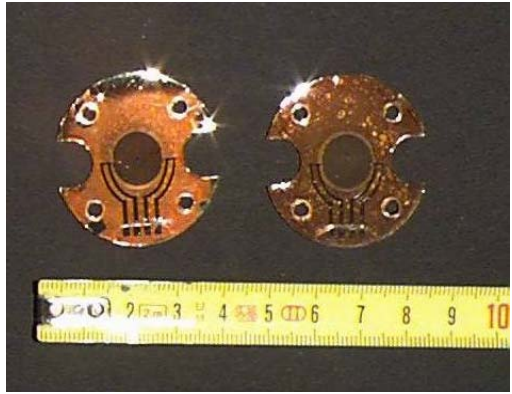

Fig. 12: Thermochimica Acta J.L. Garden

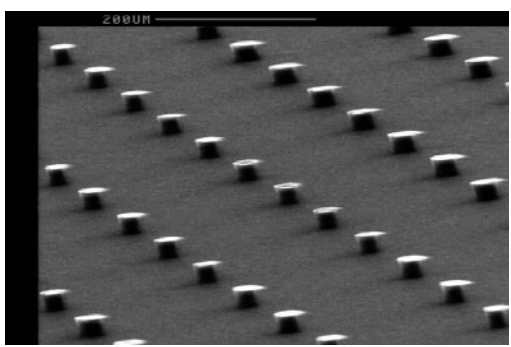

Fig.13: Thermochimica Acta J.L.Garden

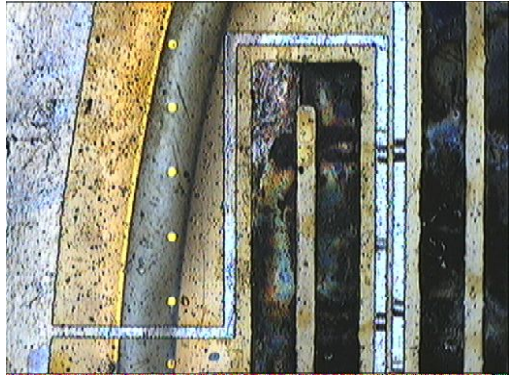

Fig. 14: Thermochimica Acta J.L.Garden.

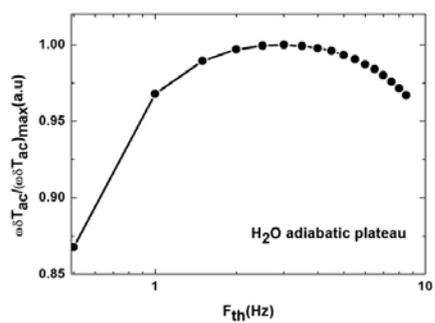

Fig.15: Thermochimica Acta J.L. Garden

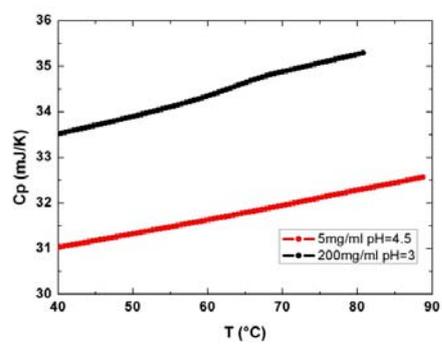

Fig.16: Thermochimica Acta J.L. Garden

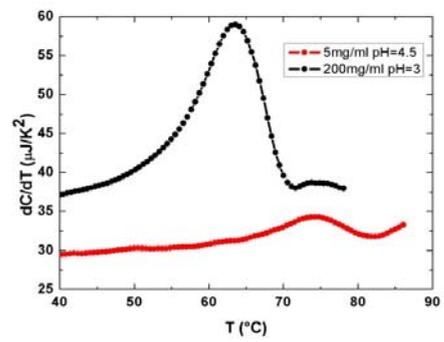

Fig. 17: Thermochimica Acta J.L. Garden

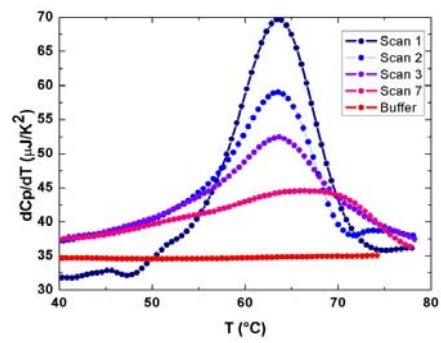

Fig. 18: Thermochimica Acta J.L.Garden

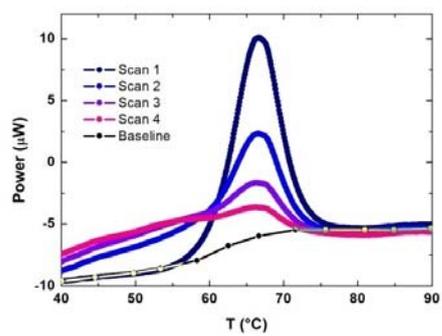

Fig. 19: Thermochimica Acta J.L. Garden

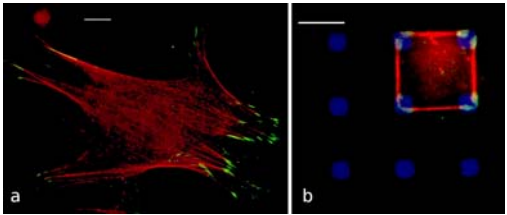

Fig. 20: Thermochimica Acta J.L. Garden

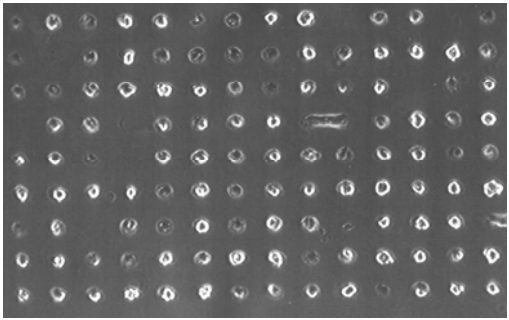

Fig. 21: Thermochimica Acta J.L. Garden

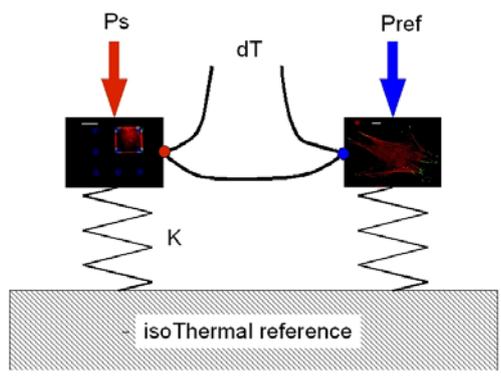

Fig.22: Thermochimica Acta J.L. Garden